\documentclass[aps,prd,amsmath,amssymb,preprintnumbers,groupedaddress,twocolumn]{revtex4-1}
\usepackage{tikz}
\usepackage{slashed}

\newcommand{\cO}{\mathcal{O}}

\newcommand{\cL}{\mathcal{L}}

\newcommand{\trh}{T_{\text{RH}}}
\newcommand{\tfo}{T_{\text{fo}}}
\newcommand{\tbbn}{T_{\text{BBN}}}
\newcommand{\gev}{\text{GeV}}
\newcommand{\mev}{\text{MeV}}
\newcommand{\tev}{\text{TeV}}

\begin{document}
\setlength{\baselineskip}{0.22in}

\preprint{NSF-KITP-13-136, MCTP-14-06}
\title{Vectorlike leptons as the tip of the dark matter iceberg}
\author{James Halverson} \affiliation{Kavli Institute for Theoretical
  Physics, University of California\\ Santa Barbara, CA 93106-4030 USA} 
\author{Nicholas Orlofsky and Aaron Pierce} \affiliation{Michigan Center for Theoretical Physics (MCTP), Department of Physics, University of Michigan \\Ann Arbor, MI 48109-1040 USA}
\date{\today}
\begin{abstract}
  A vectorlike lepton could make up a tiny fraction of the dark
  matter.  Its large $Z$-boson mediated direct detection cross section
  can compensate for the small relic abundance, giving rise to an
  interesting signal at future experiments---perhaps even the first
  one detected. We discuss how such a scenario might arise in the
  context of a simple non-thermal cosmology and investigate bounds
  from direct detection experiments and whether this scenario might be
  probed at the LHC. Searches for disappearing tracks appear
  promising.
\end{abstract}

\maketitle
\section{Introduction}
A well-motivated and minimal possibility is that dark matter
interactions in direct detection experiments are mediated by force
carriers of the Standard Model: the $Z$-boson or the Higgs boson.  The
$Z$-boson gives a spin-independent (SI) dark matter-nucleon scattering
cross section $\sigma_\text{SI} \approx 10^{-38}$ cm$^2$ for masses
$\sim 100$ GeV.  With bounds at this mass approaching $10^{-45}$
cm$^{2}$ \cite{Akerib:2013tjd,Aprile:2012nq}, it at first seems
counterproductive to consider dark matter with full strength couplings
to the $Z$.  Indeed, the canonical approach is to forbid the
$Z$-mediated operator relevant for SI scattering ${\mathcal O}_{Z}=
(\bar{q} \gamma_{\mu} q)(\bar{X} \gamma^{\mu} X)$, e.g., by making the
dark matter Majorana.  (In this case, Higgs boson exchange may yield
direct detection cross sections close to the typical bounds, see
e.g.~\cite{Cohen:2011ec,Cheung:2012qy}.)

However, it is only necessary to forbid this operator if the relic
makes up the entirety of the dark matter.  A relic may comprise a
miniscule fraction of the dark matter, but its enormous direct
detection scattering cross section can lead to an interesting signal
(See \cite{Duda:2001ae} for related work on detecting a subdominant
component of the DM in the context of the MSSM.).  We will see
dilution by the necessary amount is possible in simple cosmologies.
We briefly discuss ways in which such a particle---which could be the
first discovery at direct detection experiments---might be
disentangled from the dominant dark matter using information from both
colliders and direct detection.

  As a concrete example, consider the addition of a vectorlike pair of
  doublets $X$, $\bar{X}$ to the minimal supersymmetric standard model
  (MSSM).  What mass scale might we expect for these doublets?  Some
  mechanism must generate a mass $\mu$ for the Higgsinos of the MSSM.
  Whether this is the Giudice-Masiero mechanism \cite{Giudice:1988yz},
  the vacuum expectation value of a singlet (as in the next-to minimal
  supersymmetric standard model
  \cite{Nilles:1982dy,*Frere:1983ag,*Derendinger:1983bz}), or a
  D-brane instanton
  \cite{Blumenhagen:2006xt,*Ibanez:2006da,*Florea:2006si}, it is
  plausible that whatever generates $\mu$ would also generate a mass
  $\mu_X$ at the same scale. Naturalness arguments then indicate a
  $\mu \sim \mu_X \sim 100$ GeV -- 1 TeV.  If $X$, $\bar{X}$ have an
  unbroken $X \rightarrow -X$ symmetry, the Dirac fermion will
  comprise a component of the dark matter.  Because it is Dirac, it
  has full-strength direct detection cross section per nucleon, e.g.
  \cite{Goodman:1984dc, Essig:2007az}:
\begin{equation} \label{eqn:xsec} \sigma \approx \frac{G_{F}^{2}}{2
    \pi} \mu_{XN}^2 \frac{1}{A^2} \left[(1- 4 \sin^2 \theta_W) Z -
    (A-Z) \right]^2,
\end{equation}
where $\mu_{XN}$ is the dark matter-nucleon reduced mass.

Another potential motivation for novel vectorlike doublets arises via
string theory.  There Ref.~\cite{Halverson:2013ska} has shown that
there exist additional constraints on the chiral spectrum of SU(2)
gauge theories which ensure anomaly cancellation in nucleated D-brane
theories.  These constraints go beyond standard anomaly cancellation
in the SU(2) theories and can require the existence of electroweak
exotics; see \cite{Cvetic:2011iq,*Cvetic:2012kj} for particle physics
implications. In weakly coupled orientifold compactifications, doublet
quantum numbers for the exotics are a likely possibility.  The exotic
states arise from open strings, which selects out SU(2) singlets,
doublets, and triplets as the only possibilities. If one further
requires that one end of the open string ends on a D-brane
corresponding to a novel symmetry (perhaps related to the stability of
the $X$), then doublet quantum numbers are uniquely selected.  A string scale mass term for the doublets is forbidden by
symmetry, rendering their presence at the TeV scale even more
plausible. If the low energy theory is the Standard Model plus these
exotics, gauge couplings approximately unify at $10^{14}$ GeV
\cite{ArkaniHamed:2005yv, Mahbubani:2005pt}.

We will work under the assumption that a new Dirac SU(2) doublet $X$
exists in the 100 GeV to 1 TeV range. We address implications for
direct detection experiments and sketch how the dilution necessary to
bring it into compliance with experimental bounds might be
accomplished in a simple non-thermal cosmology.  Finally, we discuss
prospects for probing such a doublet at the LHC.
 
\section{Direct Detection and Cosmology}

We have used MicrOMEGAs \cite{Belanger:2013oya} to calculate the
spin-independent cross section of $X$ and verified that it is
consistent with Eq.~(\ref{eqn:xsec}).  We also used it to calculate
the $X$ relic abundance assuming a standard thermal freeze-out.  It is
well approximated by $\Omega_X h^2 \simeq 0.1 \left(\frac{\mu}{1~\tev}
\right)^2$ \cite{oai:arXiv.org:hep-ph/0601041}.  An $X$ produced with
a standard thermal history is well excluded by current direct
detection bounds.  To evade current bounds from LUX
\cite{Akerib:2013tjd}, $X$ with $\mu_X = 100$ GeV (1 TeV) must have a
tiny relic density $\Omega_X / \Omega_{\rm{cdm}} \lesssim 5\times
10^{-7}$ ($4\times 10^{-6}$), where $\Omega_{\rm{cdm}} h^2 = 0.1199
\pm 0.0027$ \cite{Ade:2013zuv}.

One possibility is to simply declare a smaller relic abundance by
fiat.  Indeed, we could imagine that there is thermal freeze-out with
a subsequent dilution by e.g.~late time inflation.  Interestingly,
however, the maximum dilution is limited if the baryon number is
generated before this dilution.  Even if the baryon asymmetry
proceeds by an extraordinarily efficient mechanism like Affleck-Dine
(for a review see \cite{Dine:2003ax}), where the baryon to photon
ratio could be as large as ${\mathcal O}(1)$, consistency with the
current ratio imposes a maximum dilution factor of $10^{9}$
\cite{Beringer:1900zz}.  Then, the dark matter densities would range
from $\Omega_{\rm{dil}}h^2 \approx 10^{-12} - 10^{-10}$ for $\mu_{X}=
100 - 1000$ GeV.  But given the large direct detection cross sections,
it should be possible to probe relic abundances of
$\Omega_{\rm{dil}}h^2 \approx 5 \times 10^{-11} - 10^{-11}$ without
running afoul of the neutrino background \cite{Cushman:2013zza}.  A 1
ton Xe experiment might be sensitive to relic densities perhaps a
hundred times these; therefore, it is possible to almost completely
probe this scenario of arbitrary dilution. We explore a perhaps better
motivated possibility below, where we discuss a more concrete
cosmology. In that case, the relic abundance is expected be less
diluted, and therefore the likelihood of direct detection is even
greater.

This model gives a characteristic material dependence at direct
detection experiments.  The ratio of measured cross section per
nucleon at experiments composed of Xenon, Germanium, and Argon would
be $1:0.89:0.86$.  Observing the deviation of these ratios from unity
will be challenging but would be powerful evidence for this scenario.
Also, if the mass is close to 100 GeV, it is possible to make a
determination of the $X$ mass via an examination of the recoil
spectrum, e.g., \cite{Green:2008pf}.  This mass could then be
correlated with collider discoveries, see below.

\subsection{Non-thermal production via modulus decay}
The late decay of a scalar field $\phi$ can modify the dark matter
relic abundance.  This occurs if the energy density of the universe
becomes $\phi$-dominated until the time of $\phi$ decay, which can
then both produce dark matter and provide substantial entropy
generation as it reheats the universe to a temperature $\trh$. Such
cosmologies are well-motivated in string compactifications, which
typically contain many light scalar fields in the form of stabilized
moduli.  Another possibile motivation is supersymmetric axion
models---the saxion could play the role of $\phi$ and the axion and
lightest neutralino or axino could make up (some or all of) the
remaining dark matter \cite{Baer:2011uz}.

$\trh$ is model dependent, but it is bounded by phenomenological
requirements. First, to ensure that the successful predictions of big
bang nucleosynthesis (BBN) are not spoiled, $\trh\gtrsim \tbbn \simeq
5\, \mev$ \cite{Gelmini:2006pw, *Gelmini:2006pq}. Second, to
accommodate an alternative production mechanism, it must be below the
thermal freeze-out temperature, $\trh \lesssim \tfo \simeq \mu_X/20$.
We will see that consistency with direct detection bounds will place
further limits on $\trh$.

$\trh$ is determined by the decay rate $\Gamma_\phi =
c_\phi^2\,\,m_\phi^3/M_P^2$ (perhaps arising from an operator like
$\phi G \tilde G$).  Under the assumption that $\phi$ decay and the
subsequent thermalization are instantaneous:
\begin{align}
\trh &= \left[\left(\frac{8}{90}\, \pi^3\, g_\star \right)^{-1/2} M_P \, \Gamma_\phi\right]^{1/2} \nonumber \\ &\simeq 10 \, c_\phi \,\left( \frac{m_\phi}{100\, \tev}\right)^{3/2} \,\, \mev,
\label{eqn:trh}
\end{align}
where $c_\phi$ is a presumably $\cO(1)$ constant computable in
specific models.

The $X$ relic abundance $\Omega_Xh^2$ depends on $\trh$ and $b$, the
number of dark matter particles produced per $\phi$ decay. In the
Boltzmann equations, only the combination $b/m_\phi$ appears
\cite{Gelmini:2006pw, *Gelmini:2006pq}; accordingly, we will employ
the dimensionless parameter $\eta \equiv b\, (100\, \tev/m_\phi)$. If
the $\phi$ branching ratio to $X$ is very small or zero, the $X$ relic
density is set by thermal production and freeze-out followed by its
dilution via the entropy produced in $\phi$ decays.  Since the $X$
interaction cross section is large enough to reach chemical
equilibrium prior to freeze-out, the (diluted) thermal relic density
is parametrized by \cite{McDonald:1989jd, Gelmini:2006pw,
  *Gelmini:2006pq},
\begin{equation} \label{case2}
	\Omega_X \simeq 
	\frac{\trh^3 \tfo}{(\tfo^\text{new})^4} \Omega_\text{std} \simeq
	\left(\frac{\trh}{\tfo}\right)^4 \Omega_\text{std}, \; \; (\eta \rm{\; tiny}),
\end{equation}
where $\Omega_\text{std}$ is the $X$ relic abundance assuming a
standard thermal history (i.e., $\trh > \tfo$).

On the other hand, $\eta$ may be large enough for non-thermal
production to dominate over thermal production.  Since direct
detection bounds require that the relic abundance be much less than
the standard thermal abundance, $\eta$ must nonetheless be small.
Thus, non-thermal production will not be compensated by annihilations.
In this regime \cite{Gelmini:2006pw, *Gelmini:2006pq},
\begin{equation}
  \frac{\Omega_X}{\Omega_\text{cdm}} \simeq 2\times 10^3\, \eta 
 \left( \frac{\mu_X}{100\, \gev}\right) \left(\frac{\trh}{\mev}\right), \; \; (\eta \rm{\; small}).
  \label{eqn:nonthermal dilution}
\end{equation}

\begin{figure}
\includegraphics[scale=.4]{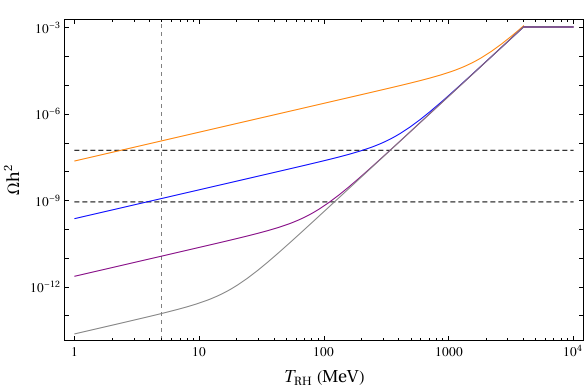}
\caption{Relic density as a function of $\trh$ for various values of
  $\eta$ for $\mu_X=100$ GeV.  The solid lines correspond to, from top
  to bottom, $\eta = 10^{-10}$ (orange), $10^{-12}$ (blue),
  $10^{-14}$ (purple), and $10^{-16}$ (gray).  The
  horizontal dashed lines correspond to the maximum allowed relic
  density to evade current LUX (top) and prospective Xe1T (bottom)
  bounds.  The vertical dashed line represents the cutoff of allowable
  $\trh$ due to BBN.}
 \label{figRelicDens}
\end{figure}

Using the sum of expressions (\ref{case2}) and (\ref{eqn:nonthermal
  dilution}) (a good approximation to the numerical solution in this
small-$\eta$ regime when $\trh$ is not too close to $\tfo$) and
assuming $\tfo \simeq \mu_X / 25$ (in good agreement with MicrOMEGAs and with numerical solutions in \cite{Gelmini:2006pw, *Gelmini:2006pq}), the relic density for $\mu_X=100$
GeV is plotted as a function of $\trh$ for various values of $\eta$ in
Figure \ref{figRelicDens}, see also Ref.~\cite{Gelmini:2006pw,
  *Gelmini:2006pq}.  Also shown are current bounds from LUX
\cite{Akerib:2013tjd} and BBN as well as prospective bounds from a
ton-scale Xe experiment \cite{Aprile:2012zx}.

Figure \ref{figRelicDens} indicates that bounds from LUX require $\eta
\lesssim 10^{-10}$. This is approximately true for any value of $\mu_X
> 100$ GeV because the LUX bound on $\sigma_\text{SI} \propto \mu_X$,
$\Omega_X \propto \eta \mu_X$ from (\ref{eqn:nonthermal dilution}),
and the cross section (\ref{eqn:xsec}) is approximately constant with
respect to $\mu_X$.

Such small values of $\eta$ require that that the Yukawa coupling
which determines $\Gamma_{\phi \rightarrow XX}$ is very small.  Is it
reasonable to expect such suppression? If $\phi$ is uncharged under
the Peccei-Quinn-like symmetry protecting $\mu_X$, then the bare
operator $\phi\, X \bar{X}$ is forbidden. However, the effect which
generates $\mu_X$ will typically also give rise to an effective Yukawa
coupling which is $\frac{\mu_X}{M_P}$ suppressed. For example, if
$\mu_X$ is generated via singlet expectation value via a coupling $s\,
X \bar{X}$, then the invariant coupling $ \frac{1}{M_P} \phi \,s X
\bar{X}$ gives an effective Yukawa $\frac{\langle s \rangle}{M_P}\,
\phi X \bar{X}$ with $\langle s \rangle \simeq \mu_X$.
\footnote{Similar statements can be made about operators arising from
  non-renormalizable Kahler potential terms e.g. $\phi^{\dagger} X
  \bar{X}$ after using the equation of motion for $X$.}  We
accordingly parameterize the effective Yukawa coupling as $c_X
\frac{\mu_X}{M_P}$, giving,
\begin{equation} \label{eqn:gamma and b}
\Gamma_{\phi XX} = \frac{m_\phi}{4 \pi} \left(c_X \,\frac{\mu_X}{M_P}\right)^2 \,\,\, 
\Rightarrow \,\,\, b= \frac{1}{2 \pi} \left(\frac{c_X}{c_\phi}\frac{\mu_X}{m_\phi}\right)^2.
\end{equation}
Here, $\Gamma_{\phi XX}$ is the width to both $X^0 \bar{X^0}$ and $X^+
X^-$.

\begin{figure}
\includegraphics[scale=.42]{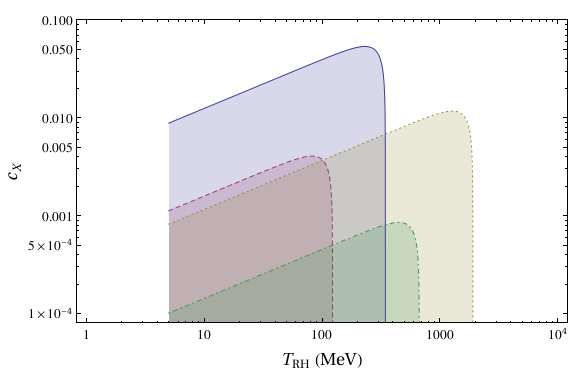}
\caption{Allowed regions (shaded) of $c_X$ for $\mu_X=100$ GeV from
  current LUX (blue solid) and prospective Xe1T (red dashed) bounds
  and for $\mu_X=1$ TeV from current LUX (yellow dotted) and
  prospective Xe1T (green dot-dashed) bounds.}
 \label{fig:c_mphi}
\end{figure}

The present constraints on $c_X$ are shown in Figure \ref{fig:c_mphi}.
Using Eqs.~(\ref{eqn:nonthermal dilution}) and (\ref{eqn:gamma and
  b}):
\begin{equation}
  \frac{\Omega_X}{\Omega_\text{cdm}} \simeq
  \frac{c_X^2}{10 \pi} \left(\frac{\mu_X}{100\, \gev}\right)^3 \left(\frac{\mev}{\trh}\right).
\end{equation}
This equation determines the upper boundary of the shaded regions.
The left boundary is set by BBN: $\trh \gtrsim \tbbn \simeq
5~\mev$. The right boundary occurs when Eq.~(\ref{case2}) is
sufficient to violate (present or future) direct detection bounds.

Absent additional model building, we expect a number of ${\mathcal
  O}(1)$ contributions to $c_{X}$, so the stringent direct detection
bounds already necessitate tuning at the few percent
level. 
The most favored $c_X$ region is the one where
the relic lies just outside current bounds in Figure~\ref{fig:c_mphi}.

The above assumes a single late-decaying modulus. However, string compactifications often contain $\cO(100)$ moduli.  For the case of many moduli, the details of the evolution of the dark matter density depend on the initial abundances of the moduli.  The single-modulus case is a good approximation to the many-moduli case under the assumption that one modulus $\phi_i$ dominates the energy density of the universe before it decays, and that no other modulus comes to dominate the energy density any time after $\phi_i$ decays.  Otherwise, several moduli may contribute to the non-thermal production of dark matter.  Additionally, decays of the later moduli might dilute contributions from earlier moduli.

\section{Collider phenomenology \label{sec:colliders}}

The strongest current collider bound on the SU(2) doublet comes from
LEP2, which bounds $\mu_X \gtrsim 95$ GeV \cite{Heister:2002mn,
  *Abbiendi:2002vz, *Abdallah:2003gv, *Acciarri:2000wy}.  We present the
prospects for 
LHC14 in the monojet and
disappearing track channels. \footnote{Ref.
  \cite{Delannoy:2013dla} studies prospects for the vector boson
  fusion (VBF) channel but concludes that it is not sensitive to a pure
  Higgsino.}

\subsection{Monojet + $\slashed{E}_\text{T}$}
The mass splitting between the $X^{\pm}$ and $X^0$ particles for mass
$\mu_X = 100$ (200) GeV is $\delta m \simeq 256$ (295) MeV.  This is a
finite, calculable effect due to electroweak symmetry breaking
\cite{Thomas:1998wy}, analogous to the charged/neutral pion mass
splitting.  With these splittings, the $X^\pm$ decay promptly into
$X^0$ + invisibly soft $\pi^\pm, ~ e^\pm \nu,$ or $\mu^\pm \nu$.
Charged and neutral doublet particles will appear as missing energy in
the detector.  The most recent results for this channel are from ATLAS
\cite{ATLAS:2012zim} and CMS \cite{CMS:rwa}.

To estimate signal and background, we use MadGraph5
\cite{Alwall:2011uj},
pass to Pythia \cite{Sjostrand:2006za} for MLM matching
\cite{Mangano:2006rw,Mrenna:2003if}, showering, and hadronization, and
use PGS \cite{PGS} for detector simulation, using an anti-$k_t$ jet
clustering algorithm with $R=.5$.
Simulated parton-level events include one or two jets.  We simulate the dominant
backgrounds $j(j) + ( Z \rightarrow \nu \nu$, $W \rightarrow l \nu,$
or $W \rightarrow \tau \nu)$, with $l=e,\mu$; the signal has $j(j) +
(X^0 \bar{X^0}, X^+ X^0, X^- \bar{X^0}$, or $X^+ X^-)$.  Following
\cite{ATLAS:2012zim}, we apply the following cuts: $(i)$
$p_\text{T}(j_1) > p_\text{T}^\text{cut}$ and $|\eta(j_1)| < 2$,
$(ii)$ $\slashed{E}_\text{T} > p_\text{T}^\text{cut}$, $(iii)$ no more
than 2 jets with $p_\text{T} > 30$ GeV and $|\eta| < 4.5$, $(iv)$
$\Delta\phi(j_2,\slashed{E}_\text{T}) > .5$, $(v)$ lepton vetoes:
$p_\text{T}(e) > 20$ GeV and $|\eta(e)| < 2.47$, $p_\text{T}(\mu) > 7$
GeV and $|\eta(\mu)| < 2.5$, or $p_\text{T}(\tau) > 20$ GeV and
$|\eta(\tau)| < 2.3$, and $(vi)$ veto on b-jets.

The significance can be found using
\begin{equation} \label{chi2} \chi^2 = \frac{S^2}{S + B +
    \sigma_{B}^2}.
\end{equation}
Its square root gives the significance.  We parameterize the
background uncertainty as $\sigma_B = \beta_\text{tot} B$, remaining
agnostic about where the uncertainties originate \footnote{Another
  approach is to estimate the use of data-driven uncertainties to
  reduce the background uncertainty as in \cite{Drees:2012dd}.  We
  found this approach tended to underestimate the uncertainties when
  applied to current experimental data \cite{ATLAS:2012zim, CMS:rwa},
  and does not substantively change the conclusions.}.
 
Taking $\cL = 3000$ fb$^{-1}$, we present significances in Table
\ref{table:monojet} assuming that $\beta_\text{tot}$ can be made to be
either .03 or .01 at $p_\text{T}^\text{cut} = 500$ GeV.  For
comparison, ATLAS and CMS have background uncertainties of
$\beta_\text{tot} \simeq .04$ and $.03$, respectively, for cuts in
present monojet analyses which have a comparable number
of background events as our projected $p_\text{T}^\text{cut}$ and
luminosity We note that \textit{if} such a small uncertainty could be
maintained at larger $p_\text{T}^\text{cut}$, the significance could
be modestly increased.

\begin{table} 
\begin{center}
\begin{tabular}{l | l l | l l l}
	& \multicolumn{2}{c |}{Signal ($\mu_X$ in GeV)} & \multicolumn{3}{c}{Backgrounds} \\
	& $\mu_X=100$ & $\mu_X=150$ & $B_{j \nu \nu}$ & $B_{j l \nu}$ & $B_{j \tau \nu}$ \\ \hline
	Cross section (fb) & 5.78 & 3.28 & 136 & 28.0 & 29.7 \\
	$\chi$ ($\beta_\text{tot}=.03$) & 1.0 & .6 & & & \\
	$\chi$ ($\beta_\text{tot}=.01$) & 3.0 & 1.7 & & & \\
\end{tabular}
\caption{Cross sections of backgrounds and signal (following cuts $(i)$ -- $(vi)$ in the text) and signal significances for the monojet + $\slashed{E}_\text{T}$ channel with $\sqrt{s}=14$ TeV, ${\mathcal L} = 3000 \rm{\; fb}^{-1}$, and $p_\text{T}^\text{cut} = 500$ GeV.  }
\label{table:monojet}
\end{center}
\end{table}

Based on our analysis, we conclude the discovery sensitivity of
this channel to a 
 SU(2) doublet appears weak.  Our signal and
background cross sections are in rough agreement with
\cite{Baer:2014cua, Han:2013usa, Schwaller:2013baa, Zhou:2013raa},
although our estimated backgrounds tend to be a little smaller and our
signals a little larger.  Thus, we reach the same general conclusions as \cite{Baer:2014cua, Han:2013usa, Schwaller:2013baa}
that $5 \sigma$ detection is unlikely in this channel, while a small
mass window may be excluded at $2 \sigma$.

\subsection{Disappearing track}
Because of the mass splitting discussed in the previous subsection,
the path length of the $X^\pm$ in its own rest frame for mass $\mu_X =
100$ (200) GeV is a modest $c \tau = 1.93$ (1.19) cm
\cite{Thomas:1998wy}.  While these path lengths are difficult to
detect, it is possible that some of the particles in the tail of the
lifetime distribution might be observed if the production rate is
sufficiently high.  Thus, low masses may be accessible to future
disappearing track studies that search for $X^\pm$ before they decay.
The most recent results from the ATLAS experiment can be found in
\cite{Aad:2013yna}.

Following the cuts in \cite{Aad:2013yna}, to obtain an estimate for
the expected signal, we use MadGraph to simulate $pp \to j + (X^+ X^0,
X^- \bar{X^0}$, or $X^+ X^-)$ at parton level, stipulating that
$p_\text{T}(j) > 90$ GeV and $|\eta(j)| < 5$.  Then, in each event
with $p_\text{T}(j_1) > p_\text{T}^\text{cut}$ (to be varied), we select the
$X^\pm$ with $.1 < |\eta^\text{track}| < 1.9$,
$p_\text{T}^\text{track} > 500$ GeV, and $p_\text{T}^\text{track} <
1000$ GeV at $\sqrt{s}=8$ TeV (to match what is done in ATLAS) and
1500 GeV at $\sqrt{s}=14$ TeV.  Next, using the known lifetime, we
calculate the probability that each passing $X^\pm$ would achieve a
transverse length of at least 30 cm before decay, corresponding to the
beginning of the first SCT layer in the ATLAS detector.  As alluded to
above, $X^\pm$ that reach the SCT are either highly boosted and/or are
in the tail of the lifetime distribution.  We assume the efficiency
for detection after these cuts is 100\%.

Comparing to current limits at $\sqrt{s}=8$ TeV, for $\mu_X=100$ GeV
and $p_\text{T}^\text{cut} = 200$ GeV, we estimate
$\sigma_\text{vis}=.27$ fb.  This is just below the ATLAS 95\%
exclusion of $\sigma_\text{vis} < .44$ fb
(smaller $p_\text{T}^\text{track}$ cuts in \cite{Aad:2013yna} set
weaker bounds).

To make projections for LHC14, we must estimate the background.  A
reliable estimate is difficult, as the dominant background (see Figure
5 of \cite{Aad:2013yna}) is from mismeasured tracks.  We parameterize
the background at $\sqrt{s}=14$ TeV and luminosity $\cL$ as
\begin{equation}
B = B_{8\, \tev}
\left(\frac{\cL}{\cL_{8\, \tev}}\right)
\left( \frac{\sigma_{14\, \tev}}{\sigma_{8\, \tev}} \right)
\left( \frac{\epsilon_{p_\text{T}(j_1) > p_\text{T}^\text{cut}}}{\epsilon_{p_\text{T}(j_1) > 90\, \gev}} \right)
P_\text{mis},
\end{equation}   
where $\cL_{8\, \tev}=20.3 ~\text{fb}^{-1}$ is the luminosity in
\cite{Aad:2013yna}, $B_{8\, \tev}$ is the estimated background in
\cite{Aad:2013yna}, $\sigma_{14\, \tev}/\sigma_{8\, \tev}$ accounts
for the increased cross section of the background as collision energy
increases (leaving all cuts constant), $\epsilon_{p_\text{T}(j_1) >
  p_\text{T}^\text{cut}}/\epsilon_{p_\text{T}(j_1) > 90\, \gev}$ accounts for a
cut intended to reduce background, and $P_\text{mis}$ parameterizes the
potential that the probability for mismeasured tracks may be greater
with increased energy and pile-up.

We obtain $B_{8\, \tev}$ by integrating the background in Figure 5 of
Ref.~\cite{Aad:2013yna} from the $p_\text{T}^\text{track}$ cut up to
1500 GeV.  We approximate $\sigma_{14\, \tev}/\sigma_{8\, \tev}
\approx 3$ based on MadGraph simulations of $p p \to j \nu \nu$, (the
dominant monojet + $\slashed{E}_\text{T}$ background).  We estimate
$\epsilon_{p_\text{T}(j_1) > p_\text{T}^\text{cut}}/\epsilon_{p_\text{T}(j_1) >
  90\, \gev}$ by applying cuts to our simulation of $pp \to j \nu
\nu$.  Finally, we assume either $P_\text{mis}=1$ or 10 and choose
$p_\text{T}(j_1)$ cuts to optimize the significance for each case.  We
underscore that many assumptions have been made to approximate $B$.
Quoted backgrounds and significances are estimates.

\begin{figure}
\includegraphics[scale=.45]{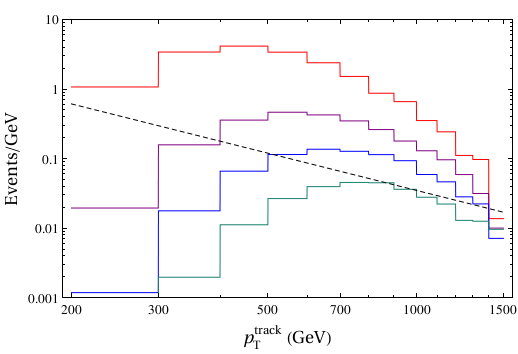}
\caption{The $p_\text{T}^\text{track}$ distribution of the background
  (black, dashed) and signal (solid) with $p_\text{T}(j_1) > 300$ GeV
  and $P_\text{mis}=1$ at $\sqrt{s}=14$ TeV and $\cL=3000$ fb$^{-1}$.
  Signal spectra correspond to, from top to bottom, $\mu_X=100$ GeV
  (red), 130 GeV (purple), 150 GeV (blue), and 170 GeV (green).}
 \label{fig:disappearingtrack}
\end{figure}

We show estimated $p_\text{T}^\text{track}$ distributions of the
background and signal at various masses for $\sqrt{s}=14$ TeV applying
the cut $p_\text{T}(j_1)>300$ GeV in Figure
\ref{fig:disappearingtrack}. For simplicity, we chose a single range
$500 ~\gev < p_\text{T}^\text{track} < 1500 ~\gev$ for all $\mu_X$.
We found optimizing this range does not affect the significance much.

The estimated backgrounds and signals for various doublet masses at $\sqrt{s} = 14$ TeV and $\cL=3000$ fb$^{-1}$ with $500~\gev < p_\text{T}^\text{track} < 1500$ GeV for different $p_\text{T}(j_1)$ cuts are shown in Table \ref{disappearingtable}.

\begin{figure}
\includegraphics[scale=.4]{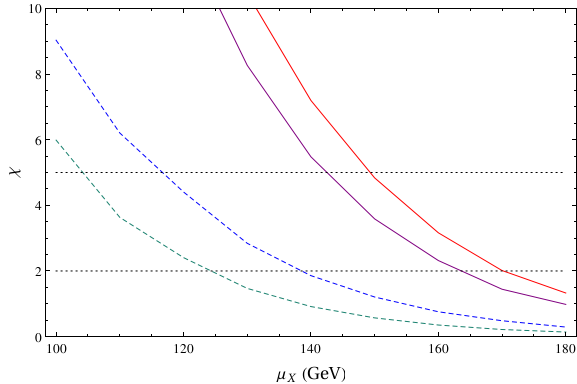}
\caption{The significance $\chi$ as a function of $\mu_X$ with
  $\sqrt{s}=14$ TeV, $500~\gev < p_\text{T}^\text{track} < 1500$ GeV, and $\sigma_B = \beta_\text{tot} B$,
  $\beta_\text{tot} =.25$.  The dashed curves are for $\cL=300$ fb$^{-1}$ assuming
  $P_\text{mis}=10$ and $p_\text{T}(j_1)>300$ GeV (green, lower) or
  $P_\text{mis}=1$ and $p_\text{T}(j_1)>200$ GeV (blue, upper).  The solid
  curves are for $\cL=3000$ fb$^{-1}$ assuming $P_\text{mis}=10$ and
  $p_\text{T}(j_1)>500$ GeV (purple, lower) or $P_\text{mis}=1$ and
  $p_\text{T}(j_1)>300$ GeV (red, upper).  All $p_T(j_1)$ cuts have been
  chosen to optimize the significance as described in the text. Dotted lines
  indicate $2\sigma$ exclusion and $5\sigma$ discovery thresholds.}
 \label{fig:disTrackSig}
\end{figure}

Significances are estimated using Eq.~(\ref{chi2}), again
parameterizing $\sigma_B = \beta_\text{tot} B$, and are shown in
Figure \ref{fig:disTrackSig} for luminosities $\cL=300$ and 3000
fb$^{-1}$, taking various $P_\text{mis}$ and optimizing the
$p_\text{T}(j_1)$ cut.  The most recent ATLAS study has cuts with
expected backgrounds of 18-48.5 events and uncertainties of about
25\%.  Naively, because a large $p_\text{T}^\text{cut}$ reduces the background
much more than the signal, a very hard $p_\text{T}^\text{cut}$ may give the best
significance.  However, because the backgrounds are estimated from
data, if the background rate is much smaller than that in the present
data, the fractional uncertainty may increase.  Thus we limit our chosen cuts to where  $B$ is roughly
in the same range as the current ATLAS backgrounds, where we assume
that the uncertainty can be approximated by $\beta_\text{tot} =.25$ \footnote{We also entertained the possibility that data-driven
  methods could decrease $\beta_\text{tot}$ as low as, e.g., .05 with
  softer cuts and larger $B$, but find harder cuts still produce
  better significance.}.  Nevertheless, an even harder cut might
ultimately be effective. 

\begin{table}
\begin{center}
\begin{tabular}{l | l | l l l l}
	& & \multicolumn{4}{c}{$S$ ($\mu_X$ in (GeV))} \\
	$p_\text{T}(j_1)$ cut & $B/P_\text{mis}$ & 100 & 130 & 150 & 170 \\ \hline
	200 GeV & 227 & 1190 & 255 & 97 & 36 \\
	300 GeV & 44.1 & 963 & 200 & 75 & 28 \\
	500 GeV & 3.96 & 646 & 137 & 49 & 18 \\
\end{tabular}
\caption{Number of the disappearing track background and signal events for $\sqrt{s}=14$ TeV, $\cL=3000$ fb$^{-1}$, and $500~\gev < p_\text{T}^\text{track} < 1500$ GeV.}
 \label{disappearingtable}
\end{center}
\end{table}

Thus, the LHC14 has the potential to probe the low-mass region of the
parameter space for an SU(2) doublet.
Optimistically, if $P_\text{mis}=1$, we estimate a $5\sigma$ discovery reach of
about $\mu_X=150$ GeV and a $2\sigma$ exclusion reach of about
$\mu_X=170$ GeV; however, a signal can still be found with larger
$P_\text{mis}$.  Further, some 
parameter space will likely be
accessible at lower luminosities.  However, the exact 
reach will depend on how the backgrounds and their
uncertainties scale with energy, instantaneous luminosity, and
$p_\text{T}^\text{cut}$.

If a signal is detected, the track length and
$p_\text{T}^\text{track}$ distributions could provide clues about the
type of particle that is detected.  For example, because of the
shortness of the $X^\pm$ lifetime, this model will have tracks with
higher $p_\text{T}$ and shorter path lengths, while other models with
longer lifetimes that have so far avoided detection will have smaller
production cross sections that compensate for the greater probability
for long path lengths.

\section{Conclusion}
We have explored the possibility that new vectorlike doublets may be
present at the TeV scale.  If stable, these particles must make up a
tiny fraction of the dark matter.  Nevertheless, they may be
phenomenologically relevant.  They could actually be the first signal
observed at direct detection experiments, perhaps presenting a
background to the true dark matter.

Comparing the two collider channels presented, the disappearing track
channel has the potential to probe a significantly larger mass range
than the monojet channel. Interestingly, it is this low mass window
where the non-thermal cosmology realizes this scenario most easily,
i.e.~with the largest values of $c_{X}$, see
Fig.~\ref{fig:c_mphi}. Observing larger-mass doublets at hadron
colliders is challenging---unless, perhaps, they are part of a larger
dark sector that boosts either the production and/or the visibility of
the events containing the $X$ particles.  If a missing energy signal
is found in the monojet channel, it will be difficult to determine
what type of particle is responsible for it, and indeed whether it
corresponds to a significant fraction of the dark matter.  A much
smaller set of models will (simultaneously) produce a disappearing
track signature.  Moreover, with enough statistics, further inferences
can in principle be made from the lifetime distribution in the
detector.

These added clues, perhaps along with material dependence at direct
detection experiments, would be enough to indicate that a putative
direct detection signal actually came from the ``tip of the dark
matter iceberg''.  A future lepton collider could also probe this
scenario contingent on kinematic accessibility.

\begin{acknowledgments}
  The authors would like to thank Lisa Everett, Jack Kearney, Sven
  Krippendorf, Scott Watson, and Daniel Whiteson. The work of JH is
  supported by the National Science Foundation under Grant
  No.~PHYS11-25915, and AP further acknowledges the KITP for support
  during his visit.  The work of NO is supported by CAREER grant
  NSF-PHY 0743315.  The work of AP is supported by DOE grant
  DE-SC0007859.
\end{acknowledgments}
\bibliography{refs2}

\begin{thebibliography}{54}%
\makeatletter
\providecommand \@ifxundefined [1]{%
 \@ifx{#1\undefined}
}%
\providecommand \@ifnum [1]{%
 \ifnum #1\expandafter \@firstoftwo
 \else \expandafter \@secondoftwo
 \fi
}%
\providecommand \@ifx [1]{%
 \ifx #1\expandafter \@firstoftwo
 \else \expandafter \@secondoftwo
 \fi
}%
\providecommand \natexlab [1]{#1}%
\providecommand \enquote  [1]{``#1''}%
\providecommand \bibnamefont  [1]{#1}%
\providecommand \bibfnamefont [1]{#1}%
\providecommand \citenamefont [1]{#1}%
\providecommand \href@noop [0]{\@secondoftwo}%
\providecommand \href [0]{\begingroup \@sanitize@url \@href}%
\providecommand \@href[1]{\@@startlink{#1}\@@href}%
\providecommand \@@href[1]{\endgroup#1\@@endlink}%
\providecommand \@sanitize@url [0]{\catcode `\\12\catcode `\$12\catcode
  `\&12\catcode `\#12\catcode `\^12\catcode `\_12\catcode `\%12\relax}%
\providecommand \@@startlink[1]{}%
\providecommand \@@endlink[0]{}%
\providecommand \url  [0]{\begingroup\@sanitize@url \@url }%
\providecommand \@url [1]{\endgroup\@href {#1}{\urlprefix }}%
\providecommand \urlprefix  [0]{URL }%
\providecommand \Eprint [0]{\href }%
\providecommand \doibase [0]{http://dx.doi.org/}%
\providecommand \selectlanguage [0]{\@gobble}%
\providecommand \bibinfo  [0]{\@secondoftwo}%
\providecommand \bibfield  [0]{\@secondoftwo}%
\providecommand \translation [1]{[#1]}%
\providecommand \BibitemOpen [0]{}%
\providecommand \bibitemStop [0]{}%
\providecommand \bibitemNoStop [0]{.\EOS\space}%
\providecommand \EOS [0]{\spacefactor3000\relax}%
\providecommand \BibitemShut  [1]{\csname bibitem#1\endcsname}%
\let\auto@bib@innerbib\@empty
\bibitem [{\citenamefont {Akerib}\ \emph {et~al.}(2013)\citenamefont {Akerib}
  \emph {et~al.}}]{Akerib:2013tjd}%
  \BibitemOpen
  \bibfield  {author} {\bibinfo {author} {\bibfnamefont {D.}~\bibnamefont
  {Akerib}} \emph {et~al.} (\bibinfo {collaboration} {LUX Collaboration}),\
  }\href@noop {} {\  (\bibinfo {year} {2013})},\ \Eprint
  {http://arxiv.org/abs/1310.8214} {arXiv:1310.8214 [astro-ph.CO]} \BibitemShut
  {NoStop}%
\bibitem [{\citenamefont {Aprile}\ \emph {et~al.}(2012)\citenamefont {Aprile}
  \emph {et~al.}}]{Aprile:2012nq}%
  \BibitemOpen
  \bibfield  {author} {\bibinfo {author} {\bibfnamefont {E.}~\bibnamefont
  {Aprile}} \emph {et~al.} (\bibinfo {collaboration} {XENON100
  Collaboration}),\ }\href {\doibase 10.1103/PhysRevLett.109.181301} {\bibfield
   {journal} {\bibinfo  {journal} {Phys.Rev.Lett.}\ }\textbf {\bibinfo {volume}
  {109}},\ \bibinfo {pages} {181301} (\bibinfo {year} {2012})},\ \Eprint
  {http://arxiv.org/abs/1207.5988} {arXiv:1207.5988 [astro-ph.CO]} \BibitemShut
  {NoStop}%
\bibitem [{\citenamefont {Cohen}\ \emph {et~al.}(2012)\citenamefont {Cohen},
  \citenamefont {Kearney}, \citenamefont {Pierce},\ and\ \citenamefont
  {Tucker-Smith}}]{Cohen:2011ec}%
  \BibitemOpen
  \bibfield  {author} {\bibinfo {author} {\bibfnamefont {T.}~\bibnamefont
  {Cohen}}, \bibinfo {author} {\bibfnamefont {J.}~\bibnamefont {Kearney}},
  \bibinfo {author} {\bibfnamefont {A.}~\bibnamefont {Pierce}}, \ and\ \bibinfo
  {author} {\bibfnamefont {D.}~\bibnamefont {Tucker-Smith}},\ }\href {\doibase
  10.1103/PhysRevD.85.075003} {\bibfield  {journal} {\bibinfo  {journal}
  {Phys.Rev.}\ }\textbf {\bibinfo {volume} {D85}},\ \bibinfo {pages} {075003}
  (\bibinfo {year} {2012})},\ \Eprint {http://arxiv.org/abs/1109.2604}
  {arXiv:1109.2604 [hep-ph]} \BibitemShut {NoStop}%
\bibitem [{\citenamefont {Cheung}\ \emph {et~al.}(2013)\citenamefont {Cheung},
  \citenamefont {Hall}, \citenamefont {Pinner},\ and\ \citenamefont
  {Ruderman}}]{Cheung:2012qy}%
  \BibitemOpen
  \bibfield  {author} {\bibinfo {author} {\bibfnamefont {C.}~\bibnamefont
  {Cheung}}, \bibinfo {author} {\bibfnamefont {L.~J.}\ \bibnamefont {Hall}},
  \bibinfo {author} {\bibfnamefont {D.}~\bibnamefont {Pinner}}, \ and\ \bibinfo
  {author} {\bibfnamefont {J.~T.}\ \bibnamefont {Ruderman}},\ }\href {\doibase
  10.1007/JHEP05(2013)100} {\bibfield  {journal} {\bibinfo  {journal} {JHEP}\
  }\textbf {\bibinfo {volume} {1305}},\ \bibinfo {pages} {100} (\bibinfo {year}
  {2013})},\ \Eprint {http://arxiv.org/abs/1211.4873} {arXiv:1211.4873
  [hep-ph]} \BibitemShut {NoStop}%
\bibitem [{\citenamefont {Duda}\ \emph {et~al.}(2002)\citenamefont {Duda},
  \citenamefont {Gelmini},\ and\ \citenamefont {Gondolo}}]{Duda:2001ae}%
  \BibitemOpen
  \bibfield  {author} {\bibinfo {author} {\bibfnamefont {G.}~\bibnamefont
  {Duda}}, \bibinfo {author} {\bibfnamefont {G.}~\bibnamefont {Gelmini}}, \
  and\ \bibinfo {author} {\bibfnamefont {P.}~\bibnamefont {Gondolo}},\ }\href
  {\doibase 10.1016/S0370-2693(02)01266-2} {\bibfield  {journal} {\bibinfo
  {journal} {Phys.Lett.}\ }\textbf {\bibinfo {volume} {B529}},\ \bibinfo
  {pages} {187} (\bibinfo {year} {2002})},\ \Eprint
  {http://arxiv.org/abs/hep-ph/0102200} {arXiv:hep-ph/0102200 [hep-ph]}
  \BibitemShut {NoStop}%
\bibitem [{\citenamefont {Giudice}\ and\ \citenamefont
  {Masiero}(1988)}]{Giudice:1988yz}%
  \BibitemOpen
  \bibfield  {author} {\bibinfo {author} {\bibfnamefont {G.}~\bibnamefont
  {Giudice}}\ and\ \bibinfo {author} {\bibfnamefont {A.}~\bibnamefont
  {Masiero}},\ }\href {\doibase 10.1016/0370-2693(88)91613-9} {\bibfield
  {journal} {\bibinfo  {journal} {Phys.Lett.}\ }\textbf {\bibinfo {volume}
  {B206}},\ \bibinfo {pages} {480} (\bibinfo {year} {1988})}\BibitemShut
  {NoStop}%
\bibitem [{\citenamefont {Nilles}\ \emph {et~al.}(1983)\citenamefont {Nilles},
  \citenamefont {Srednicki},\ and\ \citenamefont {Wyler}}]{Nilles:1982dy}%
  \BibitemOpen
  \bibfield  {author} {\bibinfo {author} {\bibfnamefont {H.~P.}\ \bibnamefont
  {Nilles}}, \bibinfo {author} {\bibfnamefont {M.}~\bibnamefont {Srednicki}}, \
  and\ \bibinfo {author} {\bibfnamefont {D.}~\bibnamefont {Wyler}},\ }\href
  {\doibase 10.1016/0370-2693(83)90460-4} {\bibfield  {journal} {\bibinfo
  {journal} {Phys.Lett.}\ }\textbf {\bibinfo {volume} {B120}},\ \bibinfo
  {pages} {346} (\bibinfo {year} {1983})}\BibitemShut {NoStop}%
\bibitem [{\citenamefont {Frere}\ \emph {et~al.}(1983)\citenamefont {Frere},
  \citenamefont {Jones},\ and\ \citenamefont {Raby}}]{Frere:1983ag}%
  \BibitemOpen
  \bibfield  {author} {\bibinfo {author} {\bibfnamefont {J.}~\bibnamefont
  {Frere}}, \bibinfo {author} {\bibfnamefont {D.}~\bibnamefont {Jones}}, \ and\
  \bibinfo {author} {\bibfnamefont {S.}~\bibnamefont {Raby}},\ }\href {\doibase
  10.1016/0550-3213(83)90606-5} {\bibfield  {journal} {\bibinfo  {journal}
  {Nucl.Phys.}\ }\textbf {\bibinfo {volume} {B222}},\ \bibinfo {pages} {11}
  (\bibinfo {year} {1983})}\BibitemShut {NoStop}%
\bibitem [{\citenamefont {Derendinger}\ and\ \citenamefont
  {Savoy}(1984)}]{Derendinger:1983bz}%
  \BibitemOpen
  \bibfield  {author} {\bibinfo {author} {\bibfnamefont {J.}~\bibnamefont
  {Derendinger}}\ and\ \bibinfo {author} {\bibfnamefont {C.~A.}\ \bibnamefont
  {Savoy}},\ }\href {\doibase 10.1016/0550-3213(84)90162-7} {\bibfield
  {journal} {\bibinfo  {journal} {Nucl.Phys.}\ }\textbf {\bibinfo {volume}
  {B237}},\ \bibinfo {pages} {307} (\bibinfo {year} {1984})}\BibitemShut
  {NoStop}%
\bibitem [{\citenamefont {Blumenhagen}\ \emph {et~al.}(2007)\citenamefont
  {Blumenhagen}, \citenamefont {Cveti{\v c}},\ and\ \citenamefont
  {Weigand}}]{Blumenhagen:2006xt}%
  \BibitemOpen
  \bibfield  {author} {\bibinfo {author} {\bibfnamefont {R.}~\bibnamefont
  {Blumenhagen}}, \bibinfo {author} {\bibfnamefont {M.}~\bibnamefont {Cveti{\v
  c}}}, \ and\ \bibinfo {author} {\bibfnamefont {T.}~\bibnamefont {Weigand}},\
  }\href {\doibase 10.1016/j.nuclphysb.2007.02.016} {\bibfield  {journal}
  {\bibinfo  {journal} {Nucl. Phys.}\ }\textbf {\bibinfo {volume} {B771}},\
  \bibinfo {pages} {113} (\bibinfo {year} {2007})},\ \Eprint
  {http://arxiv.org/abs/hep-th/0609191} {arXiv:hep-th/0609191} \BibitemShut
  {NoStop}%
\bibitem [{\citenamefont {Ibanez}\ and\ \citenamefont
  {Uranga}(2007)}]{Ibanez:2006da}%
  \BibitemOpen
  \bibfield  {author} {\bibinfo {author} {\bibfnamefont {L.}~\bibnamefont
  {Ibanez}}\ and\ \bibinfo {author} {\bibfnamefont {A.}~\bibnamefont
  {Uranga}},\ }\href {\doibase 10.1088/1126-6708/2007/03/052} {\bibfield
  {journal} {\bibinfo  {journal} {JHEP}\ }\textbf {\bibinfo {volume} {0703}},\
  \bibinfo {pages} {052} (\bibinfo {year} {2007})},\ \Eprint
  {http://arxiv.org/abs/hep-th/0609213} {arXiv:hep-th/0609213 [hep-th]}
  \BibitemShut {NoStop}%
\bibitem [{\citenamefont {Florea}\ \emph {et~al.}(2007)\citenamefont {Florea},
  \citenamefont {Kachru}, \citenamefont {McGreevy},\ and\ \citenamefont
  {Saulina}}]{Florea:2006si}%
  \BibitemOpen
  \bibfield  {author} {\bibinfo {author} {\bibfnamefont {B.}~\bibnamefont
  {Florea}}, \bibinfo {author} {\bibfnamefont {S.}~\bibnamefont {Kachru}},
  \bibinfo {author} {\bibfnamefont {J.}~\bibnamefont {McGreevy}}, \ and\
  \bibinfo {author} {\bibfnamefont {N.}~\bibnamefont {Saulina}},\ }\href
  {\doibase 10.1088/1126-6708/2007/05/024} {\bibfield  {journal} {\bibinfo
  {journal} {JHEP}\ }\textbf {\bibinfo {volume} {0705}},\ \bibinfo {pages}
  {024} (\bibinfo {year} {2007})},\ \Eprint
  {http://arxiv.org/abs/hep-th/0610003} {arXiv:hep-th/0610003 [hep-th]}
  \BibitemShut {NoStop}%
\bibitem [{\citenamefont {Goodman}\ and\ \citenamefont
  {Witten}(1985)}]{Goodman:1984dc}%
  \BibitemOpen
  \bibfield  {author} {\bibinfo {author} {\bibfnamefont {M.~W.}\ \bibnamefont
  {Goodman}}\ and\ \bibinfo {author} {\bibfnamefont {E.}~\bibnamefont
  {Witten}},\ }\href {\doibase 10.1103/PhysRevD.31.3059} {\bibfield  {journal}
  {\bibinfo  {journal} {Phys.Rev.}\ }\textbf {\bibinfo {volume} {D31}},\
  \bibinfo {pages} {3059} (\bibinfo {year} {1985})}\BibitemShut {NoStop}%
\bibitem [{\citenamefont {Essig}(2008)}]{Essig:2007az}%
  \BibitemOpen
  \bibfield  {author} {\bibinfo {author} {\bibfnamefont {R.}~\bibnamefont
  {Essig}},\ }\href {\doibase 10.1103/PhysRevD.78.015004} {\bibfield  {journal}
  {\bibinfo  {journal} {Phys.Rev.}\ }\textbf {\bibinfo {volume} {D78}},\
  \bibinfo {pages} {015004} (\bibinfo {year} {2008})},\ \Eprint
  {http://arxiv.org/abs/0710.1668} {arXiv:0710.1668 [hep-ph]} \BibitemShut
  {NoStop}%
\bibitem [{\citenamefont {Halverson}(2013)}]{Halverson:2013ska}%
  \BibitemOpen
  \bibfield  {author} {\bibinfo {author} {\bibfnamefont {J.}~\bibnamefont
  {Halverson}},\ }\href {\doibase 10.1103/PhysRevLett.111.261601} {\bibfield
  {journal} {\bibinfo  {journal} {Phys.Rev.Lett.}\ }\textbf {\bibinfo {volume}
  {111}},\ \bibinfo {pages} {261601} (\bibinfo {year} {2013})},\ \Eprint
  {http://arxiv.org/abs/1310.1091} {arXiv:1310.1091 [hep-th]} \BibitemShut
  {NoStop}%
\bibitem [{\citenamefont {Cveti{\v c}}\ \emph {et~al.}(2011)\citenamefont
  {Cveti{\v c}}, \citenamefont {Halverson},\ and\ \citenamefont
  {Langacker}}]{Cvetic:2011iq}%
  \BibitemOpen
  \bibfield  {author} {\bibinfo {author} {\bibfnamefont {M.}~\bibnamefont
  {Cveti{\v c}}}, \bibinfo {author} {\bibfnamefont {J.}~\bibnamefont
  {Halverson}}, \ and\ \bibinfo {author} {\bibfnamefont {P.}~\bibnamefont
  {Langacker}},\ }\href {\doibase 10.1007/JHEP11(2011)058} {\bibfield
  {journal} {\bibinfo  {journal} {JHEP}\ }\textbf {\bibinfo {volume} {11}},\
  \bibinfo {pages} {058} (\bibinfo {year} {2011})},\ \Eprint
  {http://arxiv.org/abs/1108.5187} {arXiv:1108.5187 [hep-ph]} \BibitemShut
  {NoStop}%
\bibitem [{\citenamefont {Cveti{\v c}}\ \emph {et~al.}(2013)\citenamefont
  {Cveti{\v c}}, \citenamefont {Halverson},\ and\ \citenamefont
  {Piragua}}]{Cvetic:2012kj}%
  \BibitemOpen
  \bibfield  {author} {\bibinfo {author} {\bibfnamefont {M.}~\bibnamefont
  {Cveti{\v c}}}, \bibinfo {author} {\bibfnamefont {J.}~\bibnamefont
  {Halverson}}, \ and\ \bibinfo {author} {\bibfnamefont {H.}~\bibnamefont
  {Piragua}},\ }\href {\doibase 10.1007/JHEP02(2013)005} {\bibfield  {journal}
  {\bibinfo  {journal} {JHEP}\ }\textbf {\bibinfo {volume} {1302}},\ \bibinfo
  {pages} {005} (\bibinfo {year} {2013})},\ \Eprint
  {http://arxiv.org/abs/1210.5245} {arXiv:1210.5245 [hep-ph]} \BibitemShut
  {NoStop}%
\bibitem [{\citenamefont {Arkani-Hamed}\ \emph {et~al.}(2005)\citenamefont
  {Arkani-Hamed}, \citenamefont {Dimopoulos},\ and\ \citenamefont
  {Kachru}}]{ArkaniHamed:2005yv}%
  \BibitemOpen
  \bibfield  {author} {\bibinfo {author} {\bibfnamefont {N.}~\bibnamefont
  {Arkani-Hamed}}, \bibinfo {author} {\bibfnamefont {S.}~\bibnamefont
  {Dimopoulos}}, \ and\ \bibinfo {author} {\bibfnamefont {S.}~\bibnamefont
  {Kachru}},\ }\href@noop {} {\  (\bibinfo {year} {2005})},\ \Eprint
  {http://arxiv.org/abs/hep-th/0501082} {arXiv:hep-th/0501082 [hep-th]}
  \BibitemShut {NoStop}%
\bibitem [{\citenamefont {Mahbubani}\ and\ \citenamefont
  {Senatore}(2006)}]{Mahbubani:2005pt}%
  \BibitemOpen
  \bibfield  {author} {\bibinfo {author} {\bibfnamefont {R.}~\bibnamefont
  {Mahbubani}}\ and\ \bibinfo {author} {\bibfnamefont {L.}~\bibnamefont
  {Senatore}},\ }\href {\doibase 10.1103/PhysRevD.73.043510} {\bibfield
  {journal} {\bibinfo  {journal} {Phys.Rev.}\ }\textbf {\bibinfo {volume}
  {D73}},\ \bibinfo {pages} {043510} (\bibinfo {year} {2006})},\ \Eprint
  {http://arxiv.org/abs/hep-ph/0510064} {arXiv:hep-ph/0510064 [hep-ph]}
  \BibitemShut {NoStop}%
\bibitem [{\citenamefont {Belanger}\ \emph {et~al.}(2013)\citenamefont
  {Belanger}, \citenamefont {Boudjema}, \citenamefont {Pukhov},\ and\
  \citenamefont {Semenov}}]{Belanger:2013oya}%
  \BibitemOpen
  \bibfield  {author} {\bibinfo {author} {\bibfnamefont {G.}~\bibnamefont
  {Belanger}}, \bibinfo {author} {\bibfnamefont {F.}~\bibnamefont {Boudjema}},
  \bibinfo {author} {\bibfnamefont {A.}~\bibnamefont {Pukhov}}, \ and\ \bibinfo
  {author} {\bibfnamefont {A.}~\bibnamefont {Semenov}},\ }\href@noop {} {\
  (\bibinfo {year} {2013})},\ \Eprint {http://arxiv.org/abs/1305.0237}
  {arXiv:1305.0237 [hep-ph]} \BibitemShut {NoStop}%
\bibitem [{\citenamefont {Arkani-Hamed}\ \emph {et~al.}(2006)\citenamefont
  {Arkani-Hamed}, \citenamefont {Delgado},\ and\ \citenamefont
  {Giudice}}]{oai:arXiv.org:hep-ph/0601041}%
  \BibitemOpen
  \bibfield  {author} {\bibinfo {author} {\bibfnamefont {N.}~\bibnamefont
  {Arkani-Hamed}}, \bibinfo {author} {\bibfnamefont {A.}~\bibnamefont
  {Delgado}}, \ and\ \bibinfo {author} {\bibfnamefont {G.}~\bibnamefont
  {Giudice}},\ }\href {\doibase 10.1016/j.nuclphysb.2006.02.010} {\bibfield
  {journal} {\bibinfo  {journal} {Nucl.Phys.}\ }\textbf {\bibinfo {volume}
  {B741}},\ \bibinfo {pages} {108} (\bibinfo {year} {2006})},\ \Eprint
  {http://arxiv.org/abs/hep-ph/0601041} {arXiv:hep-ph/0601041 [hep-ph]}
  \BibitemShut {NoStop}%
\bibitem [{\citenamefont {Ade}\ \emph {et~al.}(2013)\citenamefont {Ade} \emph
  {et~al.}}]{Ade:2013zuv}%
  \BibitemOpen
  \bibfield  {author} {\bibinfo {author} {\bibfnamefont {P.}~\bibnamefont
  {Ade}} \emph {et~al.} (\bibinfo {collaboration} {Planck Collaboration}),\
  }\href@noop {} {\  (\bibinfo {year} {2013})},\ \Eprint
  {http://arxiv.org/abs/1303.5076} {arXiv:1303.5076 [astro-ph.CO]} \BibitemShut
  {NoStop}%
\bibitem [{\citenamefont {Dine}\ and\ \citenamefont
  {Kusenko}(2003)}]{Dine:2003ax}%
  \BibitemOpen
  \bibfield  {author} {\bibinfo {author} {\bibfnamefont {M.}~\bibnamefont
  {Dine}}\ and\ \bibinfo {author} {\bibfnamefont {A.}~\bibnamefont {Kusenko}},\
  }\href {\doibase 10.1103/RevModPhys.76.1} {\bibfield  {journal} {\bibinfo
  {journal} {Rev.Mod.Phys.}\ }\textbf {\bibinfo {volume} {76}},\ \bibinfo
  {pages} {1} (\bibinfo {year} {2003})},\ \Eprint
  {http://arxiv.org/abs/hep-ph/0303065} {arXiv:hep-ph/0303065 [hep-ph]}
  \BibitemShut {NoStop}%
\bibitem [{\citenamefont {Beringer}\ \emph {et~al.}(2012)\citenamefont
  {Beringer} \emph {et~al.}}]{Beringer:1900zz}%
  \BibitemOpen
  \bibfield  {author} {\bibinfo {author} {\bibfnamefont {J.}~\bibnamefont
  {Beringer}} \emph {et~al.} (\bibinfo {collaboration} {Particle Data Group}),\
  }\href {\doibase 10.1103/PhysRevD.86.010001} {\bibfield  {journal} {\bibinfo
  {journal} {Phys.Rev.}\ }\textbf {\bibinfo {volume} {D86}},\ \bibinfo {pages}
  {010001} (\bibinfo {year} {2012})}\BibitemShut {NoStop}%
\bibitem [{\citenamefont {Cushman}\ \emph {et~al.}(2013)\citenamefont
  {Cushman}, \citenamefont {Galbiati}, \citenamefont {McKinsey}, \citenamefont
  {Robertson}, \citenamefont {Tait} \emph {et~al.}}]{Cushman:2013zza}%
  \BibitemOpen
  \bibfield  {author} {\bibinfo {author} {\bibfnamefont {P.}~\bibnamefont
  {Cushman}}, \bibinfo {author} {\bibfnamefont {C.}~\bibnamefont {Galbiati}},
  \bibinfo {author} {\bibfnamefont {D.}~\bibnamefont {McKinsey}}, \bibinfo
  {author} {\bibfnamefont {H.}~\bibnamefont {Robertson}}, \bibinfo {author}
  {\bibfnamefont {T.}~\bibnamefont {Tait}},  \emph {et~al.},\ }\href@noop {} {\
   (\bibinfo {year} {2013})},\ \Eprint {http://arxiv.org/abs/1310.8327}
  {arXiv:1310.8327 [hep-ex]} \BibitemShut {NoStop}%
\bibitem [{\citenamefont {Green}(2008)}]{Green:2008pf}%
  \BibitemOpen
  \bibfield  {author} {\bibinfo {author} {\bibfnamefont {A.~M.}\ \bibnamefont
  {Green}},\ }\href@noop {} {\bibfield  {journal} {\bibinfo  {journal} {PoS}\
  }\textbf {\bibinfo {volume} {IDM2008}},\ \bibinfo {pages} {108} (\bibinfo
  {year} {2008})},\ \Eprint {http://arxiv.org/abs/0809.1904} {arXiv:0809.1904
  [astro-ph]} \BibitemShut {NoStop}%
\bibitem [{\citenamefont {Baer}\ \emph {et~al.}(2012)\citenamefont {Baer},
  \citenamefont {Lessa},\ and\ \citenamefont {Sreethawong}}]{Baer:2011uz}%
  \BibitemOpen
  \bibfield  {author} {\bibinfo {author} {\bibfnamefont {H.}~\bibnamefont
  {Baer}}, \bibinfo {author} {\bibfnamefont {A.}~\bibnamefont {Lessa}}, \ and\
  \bibinfo {author} {\bibfnamefont {W.}~\bibnamefont {Sreethawong}},\ }\href
  {\doibase 10.1088/1475-7516/2012/01/036} {\bibfield  {journal} {\bibinfo
  {journal} {JCAP}\ }\textbf {\bibinfo {volume} {1201}},\ \bibinfo {pages}
  {036} (\bibinfo {year} {2012})},\ \Eprint {http://arxiv.org/abs/1110.2491}
  {arXiv:1110.2491 [hep-ph]} \BibitemShut {NoStop}%
\bibitem [{\citenamefont {Gelmini}\ and\ \citenamefont
  {Gondolo}(2006)}]{Gelmini:2006pw}%
  \BibitemOpen
  \bibfield  {author} {\bibinfo {author} {\bibfnamefont {G.~B.}\ \bibnamefont
  {Gelmini}}\ and\ \bibinfo {author} {\bibfnamefont {P.}~\bibnamefont
  {Gondolo}},\ }\href {\doibase 10.1103/PhysRevD.74.023510} {\bibfield
  {journal} {\bibinfo  {journal} {Phys.Rev.}\ }\textbf {\bibinfo {volume}
  {D74}},\ \bibinfo {pages} {023510} (\bibinfo {year} {2006})},\ \Eprint
  {http://arxiv.org/abs/hep-ph/0602230} {arXiv:hep-ph/0602230 [hep-ph]}
  \BibitemShut {NoStop}%
\bibitem [{\citenamefont {Gelmini}\ \emph {et~al.}(2006)\citenamefont
  {Gelmini}, \citenamefont {Gondolo}, \citenamefont {Soldatenko},\ and\
  \citenamefont {Yaguna}}]{Gelmini:2006pq}%
  \BibitemOpen
  \bibfield  {author} {\bibinfo {author} {\bibfnamefont {G.}~\bibnamefont
  {Gelmini}}, \bibinfo {author} {\bibfnamefont {P.}~\bibnamefont {Gondolo}},
  \bibinfo {author} {\bibfnamefont {A.}~\bibnamefont {Soldatenko}}, \ and\
  \bibinfo {author} {\bibfnamefont {C.~E.}\ \bibnamefont {Yaguna}},\ }\href
  {\doibase 10.1103/PhysRevD.74.083514} {\bibfield  {journal} {\bibinfo
  {journal} {Phys.Rev.}\ }\textbf {\bibinfo {volume} {D74}},\ \bibinfo {pages}
  {083514} (\bibinfo {year} {2006})},\ \Eprint
  {http://arxiv.org/abs/hep-ph/0605016} {arXiv:hep-ph/0605016 [hep-ph]}
  \BibitemShut {NoStop}%
\bibitem [{\citenamefont {McDonald}(1991)}]{McDonald:1989jd}%
  \BibitemOpen
  \bibfield  {author} {\bibinfo {author} {\bibfnamefont {J.}~\bibnamefont
  {McDonald}},\ }\href {\doibase 10.1103/PhysRevD.43.1063} {\bibfield
  {journal} {\bibinfo  {journal} {Phys.Rev.}\ }\textbf {\bibinfo {volume}
  {D43}},\ \bibinfo {pages} {1063} (\bibinfo {year} {1991})}\BibitemShut
  {NoStop}%
\bibitem [{\citenamefont {Aprile}(2012)}]{Aprile:2012zx}%
  \BibitemOpen
  \bibfield  {author} {\bibinfo {author} {\bibfnamefont {E.}~\bibnamefont
  {Aprile}} (\bibinfo {collaboration} {XENON1T collaboration}),\ }\href@noop {}
  {\  (\bibinfo {year} {2012})},\ \Eprint {http://arxiv.org/abs/1206.6288}
  {arXiv:1206.6288 [astro-ph.IM]} \BibitemShut {NoStop}%
\bibitem [{Note1()}]{Note1}%
  \BibitemOpen
  \bibinfo {note} {Similar statements can be made about operators arising from
  non-renormalizable Kahler potential terms e.g. $\phi ^{\dagger } X \protect
  \mathaccentV {bar}016{X}$ after using the equation of motion for
  $X$.}\BibitemShut {Stop}%
\bibitem [{\citenamefont {Heister}\ \emph {et~al.}(2002)\citenamefont {Heister}
  \emph {et~al.}}]{Heister:2002mn}%
  \BibitemOpen
  \bibfield  {author} {\bibinfo {author} {\bibfnamefont {A.}~\bibnamefont
  {Heister}} \emph {et~al.} (\bibinfo {collaboration} {ALEPH Collaboration}),\
  }\href {\doibase 10.1016/S0370-2693(02)01584-8} {\bibfield  {journal}
  {\bibinfo  {journal} {Phys.Lett.}\ }\textbf {\bibinfo {volume} {B533}},\
  \bibinfo {pages} {223} (\bibinfo {year} {2002})},\ \Eprint
  {http://arxiv.org/abs/hep-ex/0203020} {arXiv:hep-ex/0203020 [hep-ex]}
  \BibitemShut {NoStop}%
\bibitem [{\citenamefont {Abbiendi}\ \emph {et~al.}(2003)\citenamefont
  {Abbiendi} \emph {et~al.}}]{Abbiendi:2002vz}%
  \BibitemOpen
  \bibfield  {author} {\bibinfo {author} {\bibfnamefont {G.}~\bibnamefont
  {Abbiendi}} \emph {et~al.} (\bibinfo {collaboration} {OPAL Collaboration}),\
  }\href {\doibase 10.1140/epjc/s2003-01237-x} {\bibfield  {journal} {\bibinfo
  {journal} {Eur.Phys.J.}\ }\textbf {\bibinfo {volume} {C29}},\ \bibinfo
  {pages} {479} (\bibinfo {year} {2003})},\ \Eprint
  {http://arxiv.org/abs/hep-ex/0210043} {arXiv:hep-ex/0210043 [hep-ex]}
  \BibitemShut {NoStop}%
\bibitem [{\citenamefont {Abdallah}\ \emph {et~al.}(2004)\citenamefont
  {Abdallah} \emph {et~al.}}]{Abdallah:2003gv}%
  \BibitemOpen
  \bibfield  {author} {\bibinfo {author} {\bibfnamefont {J.}~\bibnamefont
  {Abdallah}} \emph {et~al.} (\bibinfo {collaboration} {DELPHI
  Collaboration}),\ }\href {\doibase 10.1140/epjc/s2004-01715-7} {\bibfield
  {journal} {\bibinfo  {journal} {Eur.Phys.J.}\ }\textbf {\bibinfo {volume}
  {C34}},\ \bibinfo {pages} {145} (\bibinfo {year} {2004})},\ \Eprint
  {http://arxiv.org/abs/hep-ex/0403047} {arXiv:hep-ex/0403047 [hep-ex]}
  \BibitemShut {NoStop}%
\bibitem [{\citenamefont {Acciarri}\ \emph {et~al.}(2000)\citenamefont
  {Acciarri} \emph {et~al.}}]{Acciarri:2000wy}%
  \BibitemOpen
  \bibfield  {author} {\bibinfo {author} {\bibfnamefont {M.}~\bibnamefont
  {Acciarri}} \emph {et~al.} (\bibinfo {collaboration} {L3 Collaboration}),\
  }\href {\doibase 10.1016/S0370-2693(00)00488-3} {\bibfield  {journal}
  {\bibinfo  {journal} {Phys.Lett.}\ }\textbf {\bibinfo {volume} {B482}},\
  \bibinfo {pages} {31} (\bibinfo {year} {2000})},\ \Eprint
  {http://arxiv.org/abs/hep-ex/0002043} {arXiv:hep-ex/0002043 [hep-ex]}
  \BibitemShut {NoStop}%
\bibitem [{Note2()}]{Note2}%
  \BibitemOpen
  \bibinfo {note} {Ref. \cite {Delannoy:2013dla} studies prospects for the
  vector boson fusion (VBF) channel but concludes that it is not sensitive to a
  pure Higgsino.}\BibitemShut {Stop}%
\bibitem [{\citenamefont {Thomas}\ and\ \citenamefont
  {Wells}(1998)}]{Thomas:1998wy}%
  \BibitemOpen
  \bibfield  {author} {\bibinfo {author} {\bibfnamefont {S.~D.}\ \bibnamefont
  {Thomas}}\ and\ \bibinfo {author} {\bibfnamefont {J.~D.}\ \bibnamefont
  {Wells}},\ }\href {\doibase 10.1103/PhysRevLett.81.34} {\bibfield  {journal}
  {\bibinfo  {journal} {Phys.Rev.Lett.}\ }\textbf {\bibinfo {volume} {81}},\
  \bibinfo {pages} {34} (\bibinfo {year} {1998})},\ \Eprint
  {http://arxiv.org/abs/hep-ph/9804359} {arXiv:hep-ph/9804359 [hep-ph]}
  \BibitemShut {NoStop}%
\bibitem [{ATL(2012)}]{ATLAS:2012zim}%
  \BibitemOpen
  \href@noop {} {\bibfield  {journal} {\bibinfo  {journal} {ATLAS
  Collaboration, ATLAS-CONF-2012-147, ATLAS-COM-CONF-2012-190}\ } (\bibinfo
  {year} {2012})}\BibitemShut {NoStop}%
\bibitem [{CMS(2013)}]{CMS:rwa}%
  \BibitemOpen
  \href@noop {} {\bibfield  {journal} {\bibinfo  {journal} {CMS Collaboration,
  CMS-PAS-EXO-12-048}\ } (\bibinfo {year} {2013})}\BibitemShut {NoStop}%
\bibitem [{\citenamefont {Alwall}\ \emph {et~al.}(2011)\citenamefont {Alwall},
  \citenamefont {Herquet}, \citenamefont {Maltoni}, \citenamefont {Mattelaer},\
  and\ \citenamefont {Stelzer}}]{Alwall:2011uj}%
  \BibitemOpen
  \bibfield  {author} {\bibinfo {author} {\bibfnamefont {J.}~\bibnamefont
  {Alwall}}, \bibinfo {author} {\bibfnamefont {M.}~\bibnamefont {Herquet}},
  \bibinfo {author} {\bibfnamefont {F.}~\bibnamefont {Maltoni}}, \bibinfo
  {author} {\bibfnamefont {O.}~\bibnamefont {Mattelaer}}, \ and\ \bibinfo
  {author} {\bibfnamefont {T.}~\bibnamefont {Stelzer}},\ }\href {\doibase
  10.1007/JHEP06(2011)128} {\bibfield  {journal} {\bibinfo  {journal} {JHEP}\
  }\textbf {\bibinfo {volume} {1106}},\ \bibinfo {pages} {128} (\bibinfo {year}
  {2011})},\ \Eprint {http://arxiv.org/abs/1106.0522} {arXiv:1106.0522
  [hep-ph]} \BibitemShut {NoStop}%
\bibitem [{\citenamefont {Sjostrand}\ \emph {et~al.}(2006)\citenamefont
  {Sjostrand}, \citenamefont {Mrenna},\ and\ \citenamefont
  {Skands}}]{Sjostrand:2006za}%
  \BibitemOpen
  \bibfield  {author} {\bibinfo {author} {\bibfnamefont {T.}~\bibnamefont
  {Sjostrand}}, \bibinfo {author} {\bibfnamefont {S.}~\bibnamefont {Mrenna}}, \
  and\ \bibinfo {author} {\bibfnamefont {P.~Z.}\ \bibnamefont {Skands}},\
  }\href {\doibase 10.1088/1126-6708/2006/05/026} {\bibfield  {journal}
  {\bibinfo  {journal} {JHEP}\ }\textbf {\bibinfo {volume} {0605}},\ \bibinfo
  {pages} {026} (\bibinfo {year} {2006})},\ \Eprint
  {http://arxiv.org/abs/hep-ph/0603175} {arXiv:hep-ph/0603175 [hep-ph]}
  \BibitemShut {NoStop}%
\bibitem [{\citenamefont {Mangano}\ \emph {et~al.}(2007)\citenamefont
  {Mangano}, \citenamefont {Moretti}, \citenamefont {Piccinini},\ and\
  \citenamefont {Treccani}}]{Mangano:2006rw}%
  \BibitemOpen
  \bibfield  {author} {\bibinfo {author} {\bibfnamefont {M.~L.}\ \bibnamefont
  {Mangano}}, \bibinfo {author} {\bibfnamefont {M.}~\bibnamefont {Moretti}},
  \bibinfo {author} {\bibfnamefont {F.}~\bibnamefont {Piccinini}}, \ and\
  \bibinfo {author} {\bibfnamefont {M.}~\bibnamefont {Treccani}},\ }\href
  {\doibase 10.1088/1126-6708/2007/01/013} {\bibfield  {journal} {\bibinfo
  {journal} {JHEP}\ }\textbf {\bibinfo {volume} {0701}},\ \bibinfo {pages}
  {013} (\bibinfo {year} {2007})},\ \Eprint
  {http://arxiv.org/abs/hep-ph/0611129} {arXiv:hep-ph/0611129 [hep-ph]}
  \BibitemShut {NoStop}%
\bibitem [{\citenamefont {Mrenna}\ and\ \citenamefont
  {Richardson}(2004)}]{Mrenna:2003if}%
  \BibitemOpen
  \bibfield  {author} {\bibinfo {author} {\bibfnamefont {S.}~\bibnamefont
  {Mrenna}}\ and\ \bibinfo {author} {\bibfnamefont {P.}~\bibnamefont
  {Richardson}},\ }\href {\doibase 10.1088/1126-6708/2004/05/x040} {\bibfield
  {journal} {\bibinfo  {journal} {JHEP}\ }\textbf {\bibinfo {volume} {0405}},\
  \bibinfo {pages} {040} (\bibinfo {year} {2004})},\ \Eprint
  {http://arxiv.org/abs/hep-ph/0312274} {arXiv:hep-ph/0312274 [hep-ph]}
  \BibitemShut {NoStop}%
\bibitem [{\citenamefont {Conway}\ \emph {et~al.}(2009)\citenamefont {Conway}
  \emph {et~al.}}]{PGS}%
  \BibitemOpen
  \bibfield  {author} {\bibinfo {author} {\bibfnamefont {J.}~\bibnamefont
  {Conway}} \emph {et~al.},\ }\href@noop {} {\enquote {\bibinfo {title} {{PGS}
  -- {P}retty {G}ood {S}imulation},}\ } (\bibinfo {year} {2009}),\ \Eprint
  {http://arxiv.org/abs/\url{http://physics.ucdavis.edu/~conway/research/software/pgs/pgs4-general.htm}}
  {\url{http://physics.ucdavis.edu/~conway/research/software/pgs/pgs4-general.htm}}
  \BibitemShut {NoStop}%
\bibitem [{Note3()}]{Note3}%
  \BibitemOpen
  \bibinfo {note} {Another approach is to estimate the use of data-driven
  uncertainties to reduce the background uncertainty as in \cite
  {Drees:2012dd}. We found this approach tended to underestimate the
  uncertainties when applied to current experimental data \cite {ATLAS:2012zim,
  CMS:rwa}, and does not substantively change the conclusions.}\BibitemShut
  {Stop}%
\bibitem [{\citenamefont {Baer}\ \emph {et~al.}(2014)\citenamefont {Baer},
  \citenamefont {Mustafayev},\ and\ \citenamefont {Tata}}]{Baer:2014cua}%
  \BibitemOpen
  \bibfield  {author} {\bibinfo {author} {\bibfnamefont {H.}~\bibnamefont
  {Baer}}, \bibinfo {author} {\bibfnamefont {A.}~\bibnamefont {Mustafayev}}, \
  and\ \bibinfo {author} {\bibfnamefont {X.}~\bibnamefont {Tata}},\ }\href@noop
  {} {\  (\bibinfo {year} {2014})},\ \Eprint {http://arxiv.org/abs/1401.1162}
  {arXiv:1401.1162 [hep-ph]} \BibitemShut {NoStop}%
\bibitem [{\citenamefont {Han}\ \emph {et~al.}(2013)\citenamefont {Han},
  \citenamefont {Kobakhidze}, \citenamefont {Liu}, \citenamefont {Saavedra},
  \citenamefont {Wu} \emph {et~al.}}]{Han:2013usa}%
  \BibitemOpen
  \bibfield  {author} {\bibinfo {author} {\bibfnamefont {C.}~\bibnamefont
  {Han}}, \bibinfo {author} {\bibfnamefont {A.}~\bibnamefont {Kobakhidze}},
  \bibinfo {author} {\bibfnamefont {N.}~\bibnamefont {Liu}}, \bibinfo {author}
  {\bibfnamefont {A.}~\bibnamefont {Saavedra}}, \bibinfo {author}
  {\bibfnamefont {L.}~\bibnamefont {Wu}},  \emph {et~al.},\ }\href@noop {} {\
  (\bibinfo {year} {2013})},\ \Eprint {http://arxiv.org/abs/1310.4274}
  {arXiv:1310.4274 [hep-ph]} \BibitemShut {NoStop}%
\bibitem [{\citenamefont {Schwaller}\ and\ \citenamefont
  {Zurita}(2013)}]{Schwaller:2013baa}%
  \BibitemOpen
  \bibfield  {author} {\bibinfo {author} {\bibfnamefont {P.}~\bibnamefont
  {Schwaller}}\ and\ \bibinfo {author} {\bibfnamefont {J.}~\bibnamefont
  {Zurita}},\ }\href@noop {} {\  (\bibinfo {year} {2013})},\ \Eprint
  {http://arxiv.org/abs/1312.7350} {arXiv:1312.7350 [hep-ph]} \BibitemShut
  {NoStop}%
\bibitem [{\citenamefont {Zhou}\ \emph {et~al.}(2013)\citenamefont {Zhou},
  \citenamefont {Berge}, \citenamefont {Wang}, \citenamefont {Whiteson},\ and\
  \citenamefont {Tait}}]{Zhou:2013raa}%
  \BibitemOpen
  \bibfield  {author} {\bibinfo {author} {\bibfnamefont {N.}~\bibnamefont
  {Zhou}}, \bibinfo {author} {\bibfnamefont {D.}~\bibnamefont {Berge}},
  \bibinfo {author} {\bibfnamefont {L.}~\bibnamefont {Wang}}, \bibinfo {author}
  {\bibfnamefont {D.}~\bibnamefont {Whiteson}}, \ and\ \bibinfo {author}
  {\bibfnamefont {T.}~\bibnamefont {Tait}},\ }\href@noop {} {\  (\bibinfo
  {year} {2013})},\ \Eprint {http://arxiv.org/abs/1307.5327v2 (forthcoming)}
  {arXiv:1307.5327v2 (forthcoming) [hep-ex]} \BibitemShut {NoStop}%
\bibitem [{\citenamefont {Aad}\ \emph {et~al.}(2013)\citenamefont {Aad} \emph
  {et~al.}}]{Aad:2013yna}%
  \BibitemOpen
  \bibfield  {author} {\bibinfo {author} {\bibfnamefont {G.}~\bibnamefont
  {Aad}} \emph {et~al.} (\bibinfo {collaboration} {ATLAS Collaboration}),\
  }\href {\doibase 10.1103/PhysRevD.88.112006} {\bibfield  {journal} {\bibinfo
  {journal} {Phys.Rev.}\ }\textbf {\bibinfo {volume} {D88}},\ \bibinfo {pages}
  {112006} (\bibinfo {year} {2013})},\ \Eprint {http://arxiv.org/abs/1310.3675}
  {arXiv:1310.3675 [hep-ex]} \BibitemShut {NoStop}%
\bibitem [{Note4()}]{Note4}%
  \BibitemOpen
  \bibinfo {note} {We also entertained the possibility that data-driven methods
  could decrease $\beta _\protect \text {tot}$ as low as, e.g., .05 with softer
  cuts and larger $B$, but find harder cuts still produce better
  significance.}\BibitemShut {Stop}%
\bibitem [{\citenamefont {Delannoy}\ \emph {et~al.}(2013)\citenamefont
  {Delannoy}, \citenamefont {Dutta}, \citenamefont {Gurrola}, \citenamefont
  {Johns}, \citenamefont {Kamon} \emph {et~al.}}]{Delannoy:2013dla}%
  \BibitemOpen
  \bibfield  {author} {\bibinfo {author} {\bibfnamefont {A.~G.}\ \bibnamefont
  {Delannoy}}, \bibinfo {author} {\bibfnamefont {B.}~\bibnamefont {Dutta}},
  \bibinfo {author} {\bibfnamefont {A.}~\bibnamefont {Gurrola}}, \bibinfo
  {author} {\bibfnamefont {W.}~\bibnamefont {Johns}}, \bibinfo {author}
  {\bibfnamefont {T.}~\bibnamefont {Kamon}},  \emph {et~al.},\ }\href@noop {}
  {\  (\bibinfo {year} {2013})},\ \Eprint {http://arxiv.org/abs/1308.0355}
  {arXiv:1308.0355 [hep-ph]} \BibitemShut {NoStop}%
\bibitem [{\citenamefont {Drees}\ \emph {et~al.}(2012)\citenamefont {Drees},
  \citenamefont {Hanussek},\ and\ \citenamefont {Kim}}]{Drees:2012dd}%
  \BibitemOpen
  \bibfield  {author} {\bibinfo {author} {\bibfnamefont {M.}~\bibnamefont
  {Drees}}, \bibinfo {author} {\bibfnamefont {M.}~\bibnamefont {Hanussek}}, \
  and\ \bibinfo {author} {\bibfnamefont {J.~S.}\ \bibnamefont {Kim}},\ }\href
  {\doibase 10.1103/PhysRevD.86.035024} {\bibfield  {journal} {\bibinfo
  {journal} {Phys.Rev.}\ }\textbf {\bibinfo {volume} {D86}},\ \bibinfo {pages}
  {035024} (\bibinfo {year} {2012})},\ \Eprint {http://arxiv.org/abs/1201.5714}
  {arXiv:1201.5714 [hep-ph]} \BibitemShut {NoStop}%
\end{thebibliography}%

\end{document}